\documentclass[12pt]{article}
\pdfoutput=1
\usepackage[paper=letterpaper,margin=1.0in]{geometry}
\usepackage{amsfonts,amsmath,amssymb}
\usepackage{graphicx}
\usepackage{cite}
\usepackage{hyperref}
\usepackage{amsthm}
\usepackage{esint}
\usepackage{amssymb}
\usepackage{amsmath}
\usepackage{amsthm}
\usepackage[T1]{fontenc}
\usepackage[latin9]{inputenc}
\usepackage{float}
\usepackage{units}

\newcommand{\be}{\begin{equation}}
\newcommand{\ee}{\end{equation}}
\newcommand{\bea}{\begin{eqnarray}}
\newcommand{\eea}{\end{eqnarray}}
\newcommand{\ba}{\begin{array}}
\newcommand{\ea}{\end{array}}
\newcommand{\ben}{\begin{enumerate}}
\newcommand{\een}{\end{enumerate}}
\newcommand{\bi}{\begin{itemize}}
\newcommand{\ei}{\end{itemize}}
\newcommand{\bc}{\begin{center}}
\newcommand{\ec}{\end{center}}
\newcommand{\bfig}{\begin{figure}}
\newcommand{\efig}{\end{figure}}
\newcommand{\bq}{\begin{quotation}}
\newcommand{\eq}{\end{quotation}}
\newcommand{\bt}{\begin{table}}
\newcommand{\et}{\end{table}}
\newcommand{\btab}{\begin{tabular}}
\newcommand{\etab}{\end{tabular}}
\newcommand{\bs}{\begin{slide}}
\newcommand{\es}{\end{slide}}

\begin{document}
\thispagestyle{empty}
\renewcommand{\thefootnote}{\fnsymbol{footnote}}

${}$

\vspace{2 cm}

\bc

\textbf{\LARGE Non-commutative Gravity and \\the Einstein-Van der Waals Equation of State}
\vspace{1cm}

{\large
Simon Moolman\footnote{\texttt{simon.moolman@gmail.com}}}

\vspace{0.5cm}

\textit{NITheP, School of Physics, and Mandelstam Institute for Theoretical Physics,\\
University of the Witwatersrand, Johannesburg, WITS 2050, South Africa}\\

\vspace{0.75cm}
\underline{Dated:} \today

\ec

\vspace{1.5cm}

\begin{abstract}
A calculation by Jacobson {[}1{]} strongly implies that the field
equations which describe gravity are emergent phenomena. In this paper,
the method is extended to the case of a non-commutative spacetime.
 By making use of a non-commutative version of the Raychaudhuri
equation, a new set of non-commutative Einstein equations is
derived. The results demonstrate that it is possible to use spacetime
thermodynamics to work with non-commutative gravity without the need
to vary a non-commutative action. 
\end{abstract}

\vspace{3cm}

\newpage
\renewcommand{\thefootnote}{\arabic{footnote}}
\setcounter{footnote}{0}

\section{Introduction}

In four dimensions, black hole solutions to the Einstein equations
are determined solely by mass, electric charge and angular momentum
{[}2{]}. In hindsight, this provides the first small hint that the
behaviour of black holes is in some way similar to that of classical
thermodynamics. When describing a classical gas, it is not feasible
to keep an exhaustive list of all positions and velocities of all
the gas particles. We can restrict ourselves to a few variables and
still compute meaningful physical quantities. Similarly, a star can
be described by many physical variables, but after its collapse we
are forced into using only three.

\medskip{}
\medskip{}
This similarity is, of course, nothing more than a hint. Only with
the identification of the area of a black hole with entropy and the
establishment of the four laws of black hole mechanics {[}2{]}{[}3{]},
did black hole thermodynamics become something which could be meaningfully
explored and questioned. This means, however, that an entirely valid
question would be to ask whether the analagous behaviour signifies
something deeper.
\medskip{}

\medskip{}
The evidence for spacetime as a thermodynamic system grew when, in
1995, Jacobson {[}1{]} brought the idea of black hole thermodynamics
full circle by showing that the Einstein equations can be derived
as an equation of state. The assumption that needs to be made is that
entropy and the area of causal horizons are still equal, up to some
multiplicative constant. This is not an unreasonable assumption since
entropy measures information and causal horizons hide information
from observers in space time.

\medskip{}

\medskip{}
The Einstein equations stipulate how the geodesics of spacetime bend
in response to the presence of matter. However, these field equations
need not be assumed from the outset. Simply insist that for some matter
crossing a causal horizon, $\delta Q=TdS$ holds $(\delta Q$ being
the amount of energy moving across the horizon, $T$ being the Unruh
temperature and $dS$ being the associated increase in entropy of
the universe on the other side of the causal horizon). From the only
initial assumption, $\delta Q=TdS=\eta dA$ since $dS=\eta dA$. What
this demonstrates is that you only need assume the entropy-area relation
to show that matter will bend the geodesics in a spacetime. It will
be shown below that these simple arguments will lead to the Einstein
equations. This is a thermodynamic derivation of the equations of
General Relativity and it shows that they are, in fact, an equation
of state.

\medskip{}

\medskip{}
If the Einstein equations truly are an equation of state, this begs
the question of whether or not other equations of state exist. While
it is possible to arrive at other equations of state by changing the
entropy functional $\delta S=\kappa\delta A$, this is not the approach
which will be followed in this paper. Rather, the reasoning will be
analogous to that of classical thermodynamics. Instead of making assumptions
about the thermodynamic functionals, instead we change the assumptions
about the physics of the system in question.

\medskip{}

\medskip{}
An example of this is seen in moving from the ideal gas equation

\begin{equation}
PV=nRT
\end{equation}
to the Van der Waals equation

\begin{equation}
\left(P+\frac{n^{2}a}{V^{2}}\right)\left(V-nb\right)=nRT.
\end{equation}
The ideal gas law does not allow for a system of gas particles to
interact, nor for the particles to have any size. By giving the particles
size and the ability to interact, no broad statement about thermodynamics
is made. All that is made is a change in the assumptions about the
microstructure of the thermodynamical system. Additionally, no information
about the kinetic theory of gases nor their statistical mechanics
was needed to derive the Van der Waals equation. This surprising paucity
of information shows that it might be possible to derive a new spacetime
equation of state with relatively simple assumptions about spacetime
microstructure.

\medskip{}

\medskip{}
An assumption that is tempting to make is that there is a minimum
length in spacetime. To see a simple reason why, recall that entropy
can be written as

\begin{equation}
S=k\ln\Omega
\end{equation}
where $\Omega$ is the number of states accessible to the system.
Since $\Omega$ is an integer, the value of $S$ cannot be continuous
- it can only take on certain values. Note, however, that in classical
geometry the area of a black hole is continuous and therefore the
entropy given by $S=\nicefrac{A}{4}$ is also continuous. This seems
like a contradiction but we do similar work in statistical mechanics.
We treat variables semiclassically and use them as if they were continuous,
but we know from the microscopic nature of the theory that they are
in fact quantized.

\medskip{}

\medskip{}
The above might be a simple argument but it is nonetheless forceful
and reason enough to look for different field equations of spacetime
in which area quantization is enforced from the beginning. In this
paper we will attempt to enforce the area quantization by requiring
that the spacetime coordinates obey the non-commutative relation $\left[x^{i},x^{j}\right]=i\theta^{ij}$.
Once non-commutativity is imposed, the work of Jacobson will be used
to show how a non-commutative version of the Einstein Field Equations
can be derived.

\medskip{}

\medskip{}
In section 2, Jacobson's argument is reviewed. Section 3 begins by
addressing concerns over imposing non-commutativity on spacetime and
then provides information on the mathematics needed to work with functions
on a non-commutative manifold. After this, a non-commutative version
of the Raychaudhuri equation is derived and used to repeat Jacobson's
method for the case of non-commutative manifold.

\section{Review of Jacobson's argument}

Let us briefly recapitulate the argument of {[}1{]}. Assume that there
is some accelerating observer in a spacetime, the field equations
of which are not specified at the beginning. Since the observer has
a causal horizon and there are not yet any field equations, we are
free to specify how the generators of the horizon behave when matter
crosses them. If we assume that the area of the horizon changes proportionally
to the entropy of any matter crossing the horizon, the Raychaudhuri
equation can be used to calculate the change in area of the horizon.
If this done, we are shown in {[}1{]} that the Einstein equations
come out as a result.

\medskip{}

\medskip{}
To see how the Einstein equations can be interpreted as an equation
of state, pick a point $p$ in spacetime and make the approximation
that the space around $p$ is, locally, Minkowskian. Now choose a
small patch of 2-surface which contains $p$ and call this patch $O$.

\medskip{}

\begin{figure}[H]
\begin{centering}
\includegraphics[clip,width=6cm,height=6cm]{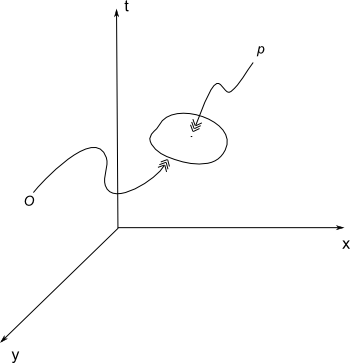}
\par\end{centering}

\protect\caption{Minkowski space containing $O$ and $p.$}
\end{figure}

\medskip{}
Choose one side of the boundary as the past of $O.$ 

\medskip{}

\begin{figure}[H]
\begin{centering}
\includegraphics[width=6cm,height=6cm]{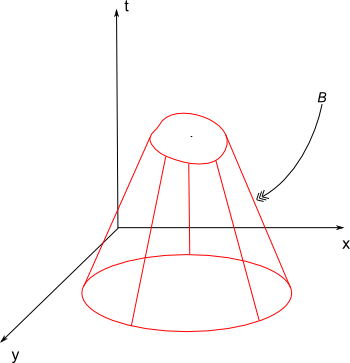}
\par\end{centering}

\protect\caption{The boundary $B$ of the past of the patch $O.$}
\end{figure}
Close to the point $p,$ this boundary is a congruence of null geodesics
orthogonal to $O$. This congruence constitutes the causal horizon
which will be studied in the derivation.

\medskip{}

\begin{figure}[H]
\begin{centering}
\includegraphics[width=6cm,height=6cm]{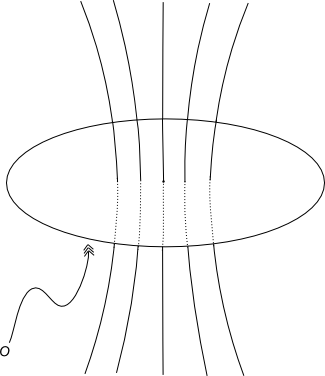}
\par\end{centering}

\protect\caption{The null geodesics orthogonal to $O$. We can use the Raychaudhuri
equation to study these geodesics and hence the behaviour of the the
patch $O.$}
\end{figure}

\medskip{}
Choose the patch so that the expansion and shear of the congruence
vanish close to $p$. It is always possible to do this. Note that
the rotation vanishes due to the fact that the congruence is normal
to the 2-surface patch. By making this construction, a local Rindler
horizon has been defined around $p$. There are local Rindler horizons
in all null directions around a spacetime point due to the fact an
observer can accelerate in any direction.

\medskip{}

\medskip{}
Since an approximately flat region of spacetime exists near our point
$p$ and around the patch $O$, the spacetime there will have all
the usual Poincare symmetries. This makes it possible to find an approximate
Killing field $\chi^{a}$ generating Lorentz boosts which are orthogonal
to $O$ and which vanish at $O.$ Suppose now that some stress energy
tensor $T_{ab}$ is defined in the spacetime. The flow of energy orthogonal
to the patch $O$ will be given by $T_{ab}\chi^{a}$.

\medskip{}

\medskip{}
Now choose $\chi^{a}$ to be future pointing to the inside past of
our patch $O$. The energy flux to the past of the patch will then
be

\begin{equation}
\delta Q=\int_{H}T_{ab}\chi^{a}d\Sigma^{b}
\end{equation}
where the integral is over the generators of the inside past horizon
of $O$. 

\medskip{}

\begin{figure}[H]
\begin{centering}
\includegraphics[width=6cm,height=6cm]{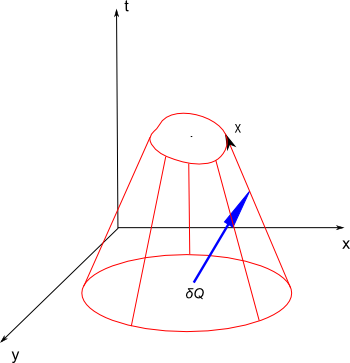}
\par\end{centering}

\protect\caption{The heat flux across the boundary $B.$ This picture illustrates that
heat (in whatever form) is moving across the causal horizon $B$ into
an area of spacetime from which a null ray can reach the Rindler observer.}
\end{figure}

\medskip{}
Call the tangent vector to the horizon generators $k^{a}$ and let
$\lambda$ be the affine parameter which vanishes at the patch $O$
and is negative to the past of $O.$ This implies 

\begin{equation}
\chi^{a}=-\kappa\lambda k^{a}
\end{equation}
and that the small patch of directed surface area is

\begin{equation}
d\Sigma^{a}=k^{a}d\lambda dA
\end{equation}
where the $dA$ is an infinitesimal piece of horizon. Rename the energy
flux to heat flux, as can be done in ordinary thermodynamics, and
write it as:

\begin{equation}
\delta Q=-\kappa\int_{H}\lambda T_{ab}k^{a}k^{b}d\lambda dA
\end{equation}
Making use of the entropy-area rule from black hole thermodynamics
makes it possible to say that the entropy of this heat flux is associated
with a small change in the area of the horizon:

\begin{equation}
dS=\eta dA.
\end{equation}
It is best to leave the constant of proportionality undetermined for
now - the rest of the derivation is insensitive to this. A small patch
of cross-sectional area of the null horizon generators is given by:

\begin{equation}
\delta A=\int_{H}\theta d\lambda dA.
\end{equation}

\begin{figure}[H]
\begin{centering}
\includegraphics[width=6cm,height=6cm]{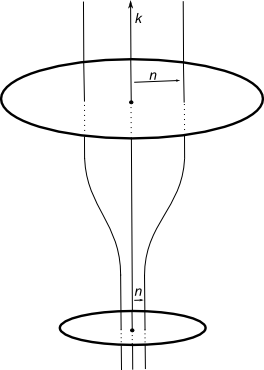}
\par\end{centering}

\protect\caption{The horizon generators expand as the entropy moves across the horizon.}
\end{figure}
Recall that the thermodynamic relations being used are:

\begin{equation}
\delta Q=TdS=\eta\delta A.
\end{equation}

Use the Unruh temperature as the temperature in the above, then set

\begin{equation}
-\kappa\int_{H}\lambda T_{ab}k^{a}k^{b}d\lambda dA=\int_{H}\theta d\lambda dA
\end{equation}
and employ the Raychaudhuri equation for null surfaces. Recall that
the congruence has been chosen to give vanishing expansion and shear,
so that the Raychaudhuri equation reduces to

\begin{equation}
\frac{d\theta}{d\lambda}=-R_{ab}k^{a}k^{b}
\end{equation}

and upon integration, $\theta$ integrates to $-\lambda R_{ab}k^{a}k^{b}.$
This tells us that (11) becomes

\begin{equation}
-\kappa\int_{H}\lambda T_{ab}k^{a}k^{b}d\lambda dA=\int_{H}-\lambda R_{ab}k^{a}k^{b}d\lambda dA
\end{equation}
and we get to

\begin{equation}
T_{ab}k^{a}k^{b}=\left(\frac{\hbar\eta}{2\pi}\right)R_{ab}k^{a}k^{b}
\end{equation}
for all null $k^{a}.$ Given $g_{ab}k^{a}k^{b}=0$ for null $k^{a},$
the metric tensor can be added into (14) for free

\begin{equation}
T_{ab}k^{a}k^{b}=\left(\frac{\hbar\eta}{2\pi}\right)R_{ab}k^{a}k^{b}+fg_{ab}k^{a}k^{b}
\end{equation}

Where $f$ is some undetermined function. Since the expressions above
are true for $any$ null vector, the result is:

\begin{equation}
T_{ab}=\left(\frac{\hbar\eta}{2\pi}\right)R_{ab}+fg_{ab}.
\end{equation}
Making use of the fact that $T_{ab}$ is divergence-free and the contracted
Bianchi identity to specify $f=-\frac{1}{2}R+\Lambda$, where $\Lambda$
is a constant, this gives {[}1{]}:

\begin{equation}
\left(\frac{2\pi}{\hbar\eta}\right)T_{ab}=R_{ab}-\frac{1}{2}Rg_{ab}+\Lambda g_{ab}
\end{equation}

\medskip{}

\medskip{}
Jacobson's calculation brings spacetime thermodynamics full circle.
By studying the behaviour of back holes, one can infer that the solutions
to the Einstein equations encode the area-entropy relationship. By
assuming that the entropy-area relationship exists, one can then get
back to the Einstein equations. More importantly, by demonstrating
that there exists a thermodynamic derivation of the equations of general
relativity, the result demonstrates that the geometrical variables
of gravity (and not just the black hole parameters) can be treated
as thermodynamic variables.

\section{The Einstein-Van der Waals Equation}

Care must be taken when it comes to making assumptions about the microstructure
of spacetime - simply placing spacetime on a lattice will break diffeomorphism
invariance. A more subtle and perhaps more useful choice is to rather
state that coordinates on spacetime do not commute:

\begin{equation}
\left[x^{i},x^{j}\right]=i\theta^{ij}.
\end{equation}
This statement has a strong analogue in quantum mechanics. By thinking
of phase space as a manifold, the quantum mechanical coordinates $\hat{x},\hat{p}$
on the phase space manifold obey the commutation relation

\begin{equation}
\left[\hat{x},\hat{p}\right]=i\hbar
\end{equation}
which demonstrates that the geometry of the phase space manifold is,
in fact, noncommutative. This setup is desirable because in both cases
it gives us cells of a definite area, but does so without forcing
either manifold onto a lattice.

\medskip{}

\medskip{}
What is desired is a way to implement these commutation relations
which still allows us to perform calculations in a straightforward
manner. The approach which will be followed in this paper is to replace
the ordinary multiplication of functions with the Moyal star product
multiplication. This star product represents the deformation of a
classical, commutative theory (the algebra of smooth functions on
the manifold) in the sense that it turns its commutative product into
a non-commutative product.

\subsection{Background}

Several steps need to be taken before we can arrive at a new set of
field equations. First, ordinary multiplication must be replaced the
Moyal star product. This is an operation which has the desirable properties
of associativity, bilinearity and the Leibniz rule. For scalars $a,b$:

\begin{equation}
(f\star g)\star h=f\star(g\star h)
\end{equation}

\begin{equation}
af\star bg=abf\star g
\end{equation}

\begin{equation}
\partial_{\mu}(f\star g)=(\partial_{\mu}f)\star g+f\star(\partial_{\mu}g).
\end{equation}

\medskip{}

\medskip{}
In order to take advantage of the above, it will be best to use the
tetrad formalism of general relativity. To set notation, we write:

\begin{equation}
g_{\mu\nu}e_{a}^{\mu}e_{b}^{\nu}=\eta_{ab}
\end{equation}

\begin{equation}
g_{\mu\nu}=e_{\mu}^{a}e_{\nu}^{b}\eta_{ab}.
\end{equation}
The Christoffel symbols and the spin connection are related by

\begin{equation}
\omega_{\mu}^{\;\; a}\;_{b}=e_{\nu}^{a}e_{b}^{\lambda}\Gamma_{\mu\lambda}^{\nu}-e_{b}^{\lambda}\partial_{\mu}e_{\lambda}^{a}.
\end{equation}
The spin connection obeys its own transformation law

\begin{equation}
\omega_{\mu}^{\;\; a'}\;_{b'}=\Lambda^{a'}\;_{a}\Lambda_{b'}\;^{b}\omega_{\mu}^{\;\; a}\;_{b}-\Lambda_{b'}\;^{c}\partial_{\mu}\Lambda^{a'}\;_{c}.
\end{equation}
which resembles the transformation law of the connection of a gauge-invariant
theory:

\begin{equation}
A_{\mu}^{\;\; A'}\;_{B'}=O^{A'}\;_{A}O_{B'}^{\; B}A_{\mu}^{\;\; A}\;_{B}-O_{B'}^{\; C}\partial_{\mu}O^{A'}\;_{C}.
\end{equation}

\medskip{}

\medskip{}
The last step in the mathematical setup is to make use of the Sieberg-Witten
map {[}4{]}. This is a way to map between a gauge theory living on
a commutative manifold and that same gauge theory living on a noncommutative
manifold. For $R^{\mu}$ with coordinates $x^{i},$ impose coordinate
noncommutativity by saying that the coordinates obey the algebra:

\[
\left[x^{i},x^{j}\right]=i\theta^{ij}
\]
where $\theta$ is real. We want to use this to deform the algebra
of functions living on $R^{n}$ to a noncommutative algebra such that

\begin{equation}
f\star g=fg+i\frac{1}{2}\theta^{ij}\partial_{i}f\partial_{j}g+O(\theta^{2}).
\end{equation}
The unique solution to this problem is {[}4{]}:

\begin{equation}
\begin{array}[t]{rl}
f(x)\star g(x) & =exp\left[\frac{i}{2}\theta^{ij}\frac{\partial}{\partial\alpha^{i}}\frac{\partial}{\partial\beta^{j}}\right]f\left(x+\alpha\right)g\left.\left(x+\beta\right)\right|_{\alpha=\beta=0}\\
 & =fg+i\frac{1}{2}\theta^{ij}\partial_{i}f\partial_{j}g+O(\theta^{2}).
\end{array}
\end{equation}
If the functions $f$ and $g$ are matrix-valued functions, then the
star product becomes the tensor product of matrix multiplication with
the star product of functions as just defined.

\medskip{}

\medskip{}
For a commutative gauge theory, the gauge transformations and field
strength are written as

\begin{equation}
\delta_{\lambda}A_{i}=\partial_{i}+i\left[\lambda,A_{i}\right],
\end{equation}

\begin{equation}
F_{ij}=\partial_{i}A_{j}-\partial_{j}A_{i}-i\left[A_{i},A_{j}\right]
\end{equation}

and

\begin{equation}
\delta_{\lambda}F_{ij}=i\left[\lambda,F_{ij}\right].
\end{equation}

For a non-commutative gauge theory, we apply the same formulae for
the gauge transformation law and the field strength, except that the
matrix multiplication is defined by the star product. If the gauge
parameter is $\hat{\lambda}$ the gauge transformations and field
strength of non-commutative Yang-Mills theory are:

\begin{equation}
\hat{\delta}_{\hat{\lambda}}\hat{A}_{i}=\partial_{i}\hat{\lambda}+i\hat{\lambda}\star\hat{A}_{i}-i\hat{A}_{i}\star\hat{\lambda}
\end{equation}

\begin{equation}
\hat{F}_{ij}=\partial_{i}\hat{A}_{j}-\partial_{j}\hat{A}_{i}-i\hat{A}_{i}\star\hat{A}_{j}+i\hat{A}_{j}\star\hat{A}_{i}
\end{equation}

\begin{equation}
\hat{\delta}_{\hat{\lambda}}\hat{F}_{ij}=i\hat{\lambda}\star\hat{F}_{ij}-i\hat{F}_{ij}\star\hat{\lambda}.
\end{equation}

To first order in $\theta,$ these expressions are:

\begin{equation}
\hat{\delta}_{\hat{\lambda}}\hat{A}_{i}=\partial_{i}\hat{\lambda}-\theta^{kl}\partial_{k}\hat{\lambda}\partial_{l}\hat{A}_{i}+O(\theta^{2})
\end{equation}

\begin{equation}
\hat{F}_{ij}=\partial_{i}\hat{A}_{j}-\partial_{j}\hat{A}_{i}-\theta^{kl}\partial_{k}\hat{A_{i}}\partial_{l}\hat{A}_{j}+O(\theta^{2})
\end{equation}

\begin{equation}
\hat{\delta}_{\hat{\lambda}}\hat{F}_{ij}=-\theta^{kl}\partial_{k}\hat{\lambda}\partial_{l}\hat{F}_{ij}+O(\theta^{2}).
\end{equation}

It now remains to find a mapping from ordinary gauge fields $A$ to
non-commutative gauge fields $\hat{A}$ which are local to any finite
order in $\theta.$ It is also necessary to impose the requirement
that if two ordinary gauge fields $A$ and $A'$ are equivalent by
an ordinary gauge transformation $U=exp(i\lambda),$ then the noncommutative
gauge fields $\hat{A}$ and $\hat{A'}$ will be gauge equivalent by
a non-commutative gauge transformation ${U=exp(i\hat{\lambda})}.$
Note that $\hat{\lambda}$ depends on $A$ and $\lambda.$ 

\medskip{}

\medskip{}
Take the gauge fields to be of rank $N$, so that the gauge parameters
are $N\times N$ matrices. It is necessary to find a mapping between
commutative and non-commutative fields such that

\begin{equation}
\hat{A}(A)+\hat{\delta}_{\hat{\lambda}}\hat{A}(A)=\hat{A}(A+\delta_{\lambda}A)
\end{equation}
where the variables $\lambda$ and $\hat{\lambda}$ are infinitesimal.
This ensures that if $A$ undergoes a transform by $\lambda,$ then
the transformation of $\hat{A}$ by $\hat{\lambda}$ is equivalent.
This forces ordinary fields which are gauge-equivalent to be mapped
to non-commutative gauge fields which are also gauge-equivalent. Working
to first order in $\theta$ and write $\hat{A=A+A'(A)}$ and $\hat{\lambda}(\lambda,A)=\lambda+\lambda'(\lambda,A).$
Thus, we expand (40) as

\begin{equation}
\begin{array}[t]{l}
A'_{i}(A+\delta_{\lambda}A)-A'_{i}(A)-\partial_{i}\lambda'-i\left[\lambda',A_{i}\right]-i\left[\lambda,A'_{i}\right]\\
=-\frac{1}{2}\theta^{kl}\left(\partial_{k}\lambda\partial_{l}A_{i}+\partial_{l}A_{i}\partial_{k}\lambda\right)+O(\theta^{2})
\end{array}
\end{equation}
where all the products appearing in the above are ordinary matrix
products. Equation (19) is solved by

\begin{equation}
\hat{A}_{i}(A)=A_{i}+A'_{i}(A)=A_{i}-\frac{1}{4}\theta^{kl}\left\{ A_{k},\partial_{l}A_{i}+F_{li}\right\} +O(\theta^{2})
\end{equation}
and

\begin{equation}
\hat{\lambda}_{i}(\lambda,A)=\lambda+\lambda'_{i}(\lambda,A)=\lambda+\frac{1}{4}\theta^{ij}\left\{ \partial_{i}\lambda,A_{j}\right\} +O(\theta^{2}).
\end{equation}
The gauge strength is then written as

\begin{equation}
\hat{F}_{ij}=F_{ij}+\frac{1}{4}\theta^{kl}\left(2\left\{ F_{ik},F_{jl}\right\} -\left\{ A_{k},D_{l}F_{ij}+\partial_{l}F_{ij}\right\} \right)+O(\theta^{2}).
\end{equation}
Equations (42), (43) and (44) illustrate what a gauge theory would
look like if it were to live on a noncommutative manifold and they
do so in terms of what the field looks like on a regular, commutative
manifold. There now exists a new noncommutative theory expressed entirely
in terms of expressions and functions we already know.

\medskip{}

\medskip{}
So far, the work has only been done to first order. To work to higher
orders in $\theta,$ consider mapping the field $\hat{A}(\theta)$
to $\hat{A}(\theta+\delta\theta).$ The only property of the $\star-$product
that one needs to check in order to see that (43) and (44) satisfy
(40) is

\begin{equation}
\delta\theta^{ij}\frac{\partial}{\partial\theta^{ij}}(f\star g)=\frac{i}{2}\delta\theta^{ij}\frac{\partial f}{\partial x^{i}}\star\frac{\partial g}{\partial x^{j}}
\end{equation}
when $\theta=0.$ Since this is true for any value of $\theta,$ it
is possible to write down formulas which tell us how $\hat{A}(\theta)$
and $\hat{\lambda}(\theta)$ change when $\theta$ is varied. These
are:

\begin{equation}
\begin{array}[t]{rl}
\delta\hat{A}_{i}(\theta) & =\delta\theta^{kl}\frac{\partial}{\partial\theta^{kl}}\hat{A}_{i}(\theta)\\
 & =-\frac{1}{4}\delta\theta^{kl}\left[\hat{A}_{k}\star\left(\partial_{l}\hat{A}_{i}+\hat{F}_{li}\right)+\left(\partial_{l}\hat{A}_{i}+\hat{F}_{li}\right)\star\hat{A_{k}}\right]
\end{array}
\end{equation}

\begin{equation}
\begin{array}[t]{rl}
\delta\hat{\lambda}(\theta) & =\delta\theta^{kl}\frac{\partial}{\partial\theta^{kl}}\hat{\lambda}(\theta)\\
 & =-\frac{1}{4}\delta\theta^{kl}\left(\partial_{k}\lambda\star A_{l}+A_{l}\star\partial_{k}\lambda\right)
\end{array}
\end{equation}
and

\begin{equation}
\begin{array}[t]{rl}
\delta\hat{F}_{ij}(\theta) & =\delta\theta^{kl}\frac{\partial}{\partial\theta^{kl}}\hat{F}_{ij}(\theta)\\
 & =-\frac{1}{4}\delta\theta^{kl}\left[2\hat{F}_{ik}\star\hat{F}_{jl}+2\hat{F}_{jl}\star\hat{F}_{ik}\right.\\
 & -\hat{A}_{k}\star\left(\hat{D}_{l}\hat{F}_{ij}+\partial_{l}\hat{F}_{ij}\right)-\left.\left(\hat{D}_{l}\hat{F}_{ij}+\partial_{l}\hat{F}_{ij}\right)\star\hat{A}_{k}\right].
\end{array}
\end{equation}

\medskip{}

\medskip{}
Work by Chamseddine {[}5{]} shows how the Sieberg-Witten map can transform
the quantities in the tetrad formalism into the same quantities living
on a noncommutative manifold. The key lies in introducing the gauge
fields $\hat{\omega}_{\mu}^{AB}$ which are subject to:

\begin{equation}
\hat{\omega}_{\mu}^{AB\dagger}(x,\theta)=-\hat{\omega}_{\mu}^{AB}(x,\theta)
\end{equation}

\begin{equation}
\hat{\omega}_{\mu}^{AB}(x,\theta)^{r}\equiv\hat{\omega}_{\mu}^{AB}(x,-\theta)=-\hat{\omega}_{\mu}^{AB}(x,\theta).
\end{equation}

Expanding these fields in terms of $\theta:$

\begin{equation}
\hat{\omega}_{\mu}^{AB}(x,\theta)=\omega_{\mu}^{AB}-i\theta^{\nu\rho}\omega_{\mu\nu\rho}^{AB}+\ldots
\end{equation}

These new fields are related to the old fields via the Sieberg-Witten
map:

\begin{equation}
\hat{\omega}_{\mu}^{AB}(\omega)+\delta_{\hat{\lambda}}\hat{\omega}_{\mu}^{AB}(\omega)=\hat{\omega}_{\mu}^{AB}(\omega+\delta_{\lambda}\omega)
\end{equation}

where $\hat{g}=e^{\hat{\lambda}}$ and the infinitesimal transformation
of of $\omega_{\mu}^{AB}$ is given by

\begin{equation}
\delta_{\lambda}\omega_{\mu}^{AB}=\partial_{\mu}\lambda^{AB}+\omega_{\mu}^{AC}\lambda^{CB}-\lambda^{AC}\omega_{\mu}^{CB}.
\end{equation}

The deformed fields are given by the same expression, but with matrix
multiplication replaced by the star product:

\begin{equation}
\delta_{\hat{\lambda}}\hat{\omega}_{\mu}^{AB}=\partial_{\mu}\hat{\lambda}^{AB}+\hat{\omega}_{\mu}^{AC}\star\hat{\lambda}^{CB}-\hat{\lambda}^{AC}\star\hat{\omega}_{\mu}^{CB}.
\end{equation}

The above equation can be solved to all orders in $\theta$ and the
result is {[}5{]}:

\begin{equation}
\delta\hat{\omega}_{\mu}^{AB}=\frac{i}{4}\theta^{\nu\rho}\left\{ \hat{\omega}_{\nu},_{\star}\partial_{\rho}\hat{\omega}_{\mu}+\hat{R}_{\rho\mu}\right\} ^{AB}.
\end{equation}
This result is necessary because it will be used to find the higher-order
corrections to the deformed tetrads. Even though it is now possible
to expand and solve the above equation, we have not yet determined
how $\hat{e}_{\mu}^{a}$ is related to the undeformed field, since
it is not a gauge field. To proceed, treat $\hat{e}_{\mu}^{a}$ as
the gauge field of the translational generator of the inhomogeneous
Lorentz group obtained by contracting $SO(4,1)$ to $ISO(3,1)$. To
do this, define the $SO(4,1)$ gauge field $\omega_{\mu}^{AB}$ with
the strength 

\begin{equation}
R_{\mu\nu}^{AB}=\partial_{\mu}\omega_{\nu}^{AB}+\partial_{\nu}\omega_{\mu}^{AB}+\omega_{\mu}^{AC}\omega_{\nu}^{CB}+\omega_{\nu}^{AC}\omega_{\mu}^{CB}
\end{equation}
where $A=a,5.$ Now define $\omega_{\mu}^{a5}=ke_{\mu}^{a}$ so that
we get

\begin{equation}
R_{\mu\nu}^{ab}=\partial_{\mu}\omega_{\nu}^{ab}-\partial_{\nu}\omega_{\mu}^{ab}+\omega_{\mu}^{ac}\omega_{\nu}^{cb}+k^{2}\left(e_{\mu}^{a}e_{\nu}^{b}-e_{\nu}^{a}e_{\mu}^{b}\right)
\end{equation}
and

\begin{equation}
R_{\mu\nu}^{a5}=kT_{\mu\nu}^{a}=k\left(\partial_{\mu}e_{\nu}^{a}-\partial_{\nu}e_{\mu}^{a}+\omega_{\mu}^{ac}e_{\nu}^{c}-\omega_{\nu}^{ac}e_{\mu}^{c}\right).
\end{equation}
Now perform the contraction by taking $k\rightarrow0$ and impose
the condition $T_{\mu\nu}^{a}=0$ so that $\omega_{\mu}^{ab}$ can
be solved for in terms of $e_{\mu}^{a}.$ 

\medskip{}

\medskip{}
For the deformed case, write $\hat{\omega}_{\mu}^{a5}=ke_{\mu}^{a}$
and $\hat{\omega}_{\mu}^{55}=k\hat{\phi_{\mu}}.$ It is not necessary
impose $\hat{T}_{\mu\nu}^{a}=0$ because $\phi_{u}$ drops out when
$k\rightarrow0.$ This gives the deformed tetrad to second order as
{[}5{]}

\begin{equation}
\begin{array}{rl}
\hat{e}_{\mu}^{a}= & \;\; e_{\mu}^{a}-\frac{i}{4}\theta^{\nu\rho}\left(\omega_{\mu}^{ac}\partial_{\rho}e_{\mu}^{c}+\left(\partial_{\rho}\omega_{\mu}^{ac}+R_{\rho\mu}^{ac}\right)e_{\nu}^{c}\right)\\
 & +\frac{1}{32}\theta^{\nu\rho}\theta^{\kappa\sigma}\left(2\left\{ R_{\sigma\nu},R_{\mu\rho}\right\} ^{ac}e_{\kappa}^{c}-\omega_{\kappa}^{ac}\left(D_{\rho}R_{\sigma\mu}^{cd}+\partial_{\rho}R_{\sigma\mu}^{cd}\right)e_{\nu}^{d}\right.\\
 & -\left\{ \omega_{\nu},\left(D_{\rho}R_{\sigma\mu}+\partial_{\rho}R_{\sigma\mu}\right)\right\} ^{ad}e_{\kappa}^{d}-\partial_{\sigma}\left\{ \omega_{\nu},\left(\partial_{\rho}\omega_{\mu}+R_{\rho\mu}\right)\right\} ^{ac}e_{\kappa}^{c}\\
 & -\omega_{\kappa}^{ac}\partial_{\sigma}\left(\omega_{\nu}^{cd}\partial_{\rho}e_{\mu}^{d}+\left(\partial_{\rho}\omega_{\mu}^{cd}+R_{\rho\mu}^{cd}\right)e_{\nu}^{d}\right)+\partial_{\nu}\omega_{\kappa}^{ac}\partial_{\rho}\partial_{\sigma}e_{\mu}^{c}\\
 & -\partial_{\rho}\left(\partial_{\sigma}\omega_{\mu}^{ac}+R_{\sigma\mu}^{ac}\right)\partial_{nu}e_{\kappa}^{c}-\left\{ \omega_{\nu},\left(\partial_{\rho}\omega_{\kappa}+R_{\rho\kappa}\right)\right\} ^{ac}\partial_{\sigma}e_{\mu}^{c}\\
 & -\left(\partial_{\sigma}\omega_{\mu}^{ac}+R_{\sigma\mu}^{ac}\right)\left.\left(\omega_{\nu}^{cd}\partial_{\rho}e_{\kappa}^{d}+\left(\partial_{\rho}\omega_{\kappa}^{cd}+R_{\rho\kappa}^{cd}\right)e_{\nu}^{d}\right)\right).
\end{array}
\end{equation}
There now exists a deformed spin connection entirely in terms of normal
geometric variables as well as a deformed tetrad entirely in terms
of normal geometric variables. These more basic indexed objects can
be used in calculations to build up more complex tensor objects. For
example, it can be shown that the expansion of $\hat{R}_{\mu\nu}^{AB}$
in terms of $\theta$

\begin{equation}
\hat{R}_{\mu\nu}^{ab}=R_{\mu\nu}^{ab}+i\theta^{\rho\tau}R_{\mu\nu\rho\tau}^{ab}+\theta^{\rho\tau}\theta^{\kappa\sigma}R_{\mu\nu\rho\tau\kappa\sigma}^{ab}.
\end{equation}
has coefficients {[}5{]}

\begin{equation}
\begin{array}{rl}
R_{\mu\nu}^{ab}= & \partial_{\mu}\omega_{\nu}^{ab}+\partial_{\nu}\omega_{\mu}^{ab}+\omega_{\mu}^{ac}\omega_{\nu}^{cb}+\omega_{\nu}^{ac}\omega_{\mu}^{cb}\\
R_{\mu\nu\rho\tau}^{ab}= & \partial_{\mu}\omega_{\nu\rho\tau}^{ab}+\omega_{\mu}^{ac}\omega_{\nu\rho\tau}^{cb}+\omega_{\mu\rho\tau}^{ac}\omega_{\nu}^{cb}-\frac{1}{2}\partial_{\rho}\omega_{\mu}^{ac}\partial_{\tau}\omega_{\nu}^{cb}-\mu\leftrightarrow\nu\\
R_{\mu\nu\rho\tau\kappa\sigma}^{ab}= & \partial_{\mu}\omega_{\nu\rho\tau\kappa\sigma}^{ab}+\omega_{\mu}^{ac}\omega_{\nu\rho\tau\kappa\sigma}^{cb}+\omega_{\mu\rho\tau\kappa\sigma}^{ac}\omega_{\nu}^{cb}-\omega_{\mu\rho\tau}^{ac}\omega_{\nu\kappa\sigma}^{cb}\\
 & -\frac{1}{4}\partial_{\rho}\partial_{\kappa}\omega_{\mu}^{ac}\partial_{\tau}\partial_{\sigma}\omega_{\nu}^{cb}-\mu\leftrightarrow\nu.
\end{array}
\end{equation}

These tetrad calculations contribute only indirectly to the derivation
of a new equation of state. It is necessary to move back to the tensors
of normal general relativity and the Raychaudhuri equation to derive
a new equation of state.

\subsection{Implementation}

The first step to take is to find an expression for the deformed Riemann
tensor which appears in the Raychaudhuri equation. Equation (60) will
not suffice because it mixes tetrad indices with coordinate-frame
indices whereas a Riemann tensor with only coordinate-frame indices
is necessary. The classical expression

\begin{equation}
R_{\mu\nu\rho}^{\sigma}=\partial_{\nu}\Gamma_{\mu\rho}^{\sigma}-\partial_{\rho}\Gamma_{\mu\nu}^{\sigma}+\Gamma_{\alpha\nu}^{\sigma}\Gamma_{\mu\rho}^{\alpha}-\Gamma_{\alpha\rho}^{\sigma}\Gamma_{\mu\nu}^{\alpha}
\end{equation}
can be replaced with

\begin{equation}
\hat{R}_{\mu\nu\rho}^{\sigma}=\partial_{\nu}\hat{\Gamma}_{\mu\rho}^{\sigma}-\partial_{\rho}\hat{\Gamma}_{\mu\nu}^{\sigma}+\hat{\Gamma}_{\alpha\nu}^{\sigma}\star\hat{\Gamma}_{\mu\rho}^{\alpha}-\hat{\Gamma}_{\alpha\rho}^{\sigma}\star\hat{\Gamma}_{\mu\nu}^{\alpha}
\end{equation}
and then expanded to higher orders in the noncommutative parameter.
To accomplish this, note that we take the classical expression

\begin{equation}
\Gamma_{\mu\lambda}^{\nu}=e_{a}^{\nu}\partial_{\mu}e_{\lambda}^{a}+e_{a}^{\nu}e_{\lambda}^{b}\omega_{\mu b}^{a}
\end{equation}
and turn it into the deformed relation

\begin{equation}
\hat{\Gamma}_{\mu\lambda}^{\nu}=\hat{e}_{a}^{\nu}\star\partial_{\mu}\hat{e}_{\lambda}^{a}+\hat{e}_{a}^{\nu}\star\hat{e}_{\lambda}^{b}\star\hat{\omega}_{\mu b}^{a}.
\end{equation}
Expressions already exist for $\hat{e}_{a}^{\nu}$ and $\hat{\omega}_{\mu b}^{a}$
so these can be used to find the second-order corrections to the Christoffel
symbols. Begin by expanding to second order:

\begin{equation}
\hat{\Gamma}_{\mu\lambda}^{\nu}=\Gamma_{\mu\lambda}^{\nu}-i\theta^{xy}\Gamma_{\mu\lambda xy}^{\nu}+\theta^{xy}\theta^{pq}\Gamma_{\mu\lambda xypq}^{\nu}.
\end{equation}
By explicitly calculating the terms in (65) just like was done for
(60), the results up to second order are

\begin{equation}
\Gamma_{\mu\lambda}^{\nu}=e_{a}^{\nu}\partial_{\mu}e_{\lambda}^{a}+e_{a}^{\nu}e_{\lambda}^{b}\omega_{\mu b}^{a},
\end{equation}

\begin{equation}
\Gamma_{\mu\lambda xy}^{\nu}=\begin{array}{cc}
-e_{a}^{\nu}\partial_{\mu}e_{\lambda xy}^{a} & -e_{axy}^{\nu}\partial_{\mu}e_{\lambda}^{a}\\
-\frac{1}{2}\partial_{x}e_{a}^{\nu}\partial_{y}\partial_{\mu}e_{\nu}^{a} & -e_{\lambda xy}^{b}e_{a}^{\nu}\omega_{\mu b}^{a}\\
+e_{axy}^{\nu}e_{\lambda}^{b}\omega_{\mu b}^{a} & \frac{1}{2}\partial_{x}e_{a}^{\nu}\partial_{y}e_{\lambda}^{b}\omega_{\mu b}^{a}\\
+e_{a}^{\nu}e_{\lambda}^{b}\omega_{\mu bxy}^{a} & -\partial_{x}e_{a}^{v}e_{\lambda}^{b}\partial_{y}\omega_{\mu b}^{a}
\end{array},
\end{equation}
and

\begin{equation}
\Gamma_{\mu\lambda xypq}^{\nu}=\begin{array}{cc}
+e_{a}^{v}\partial_{\mu}e_{\lambda xypq}^{a} & -e_{axy}^{v}\partial_{\mu}e_{\lambda pq}^{a}\\
+e_{axypq}^{v}\partial_{\mu}e_{\lambda}^{a} & -\frac{1}{2}\partial_{x}e_{a}^{v}\partial_{y}\partial_{\mu}e_{\nu pq}^{a}\\
-\frac{1}{2}\partial_{x}e_{apq}^{\nu}\partial_{y}\partial_{\mu}e_{\nu}^{a} & -\frac{1}{4}\partial_{x}\partial_{p}e_{a}^{v}\partial_{y}\partial_{q}\partial_{\mu}e_{\nu}^{a}\\
+e_{\lambda xypq}^{b}e_{a}^{v}\omega_{\mu b}^{a} & -e_{axy}^{\nu}e_{\lambda pq}^{b}\omega_{\mu b}^{a}\\
+e_{axypq}^{\nu}e_{\lambda}^{b}\omega_{\mu b}^{a} & -\frac{1}{2}\partial_{y}e_{\lambda pq}^{b}\partial_{x}e_{a}^{\nu}\omega_{\mu b}^{a}\\
-\frac{1}{4}\partial_{x}\partial_{p}e_{a}^{\nu}\partial_{y}\partial_{q}e_{\lambda}^{b}\omega_{\mu b}^{a} & +e_{\lambda xy}^{b}e_{a}^{\nu}\omega_{\mu bpq}^{a}\\
+e_{axy}^{\nu}e_{\lambda}^{b}\omega_{\mu bpq}^{a} & +\frac{1}{2}\partial_{x}e_{a}^{\nu}\partial_{y}e_{\lambda}^{b}\omega_{\mu bpq}^{a}\\
+e_{a}^{\nu}e_{\lambda}^{b}\omega_{\mu bxypq}^{a} & -\frac{1}{2}\partial_{x}e_{\lambda pq}^{b}e_{a}^{\nu}\partial_{y}\omega_{\mu b}^{a}\\
-\frac{1}{2}\partial_{x}e_{apq}^{\nu}e_{\lambda}^{b}\partial_{y}\omega_{\mu b}^{a} & -\frac{1}{4}\partial_{x}\partial_{p}e_{a}^{\nu}\partial_{q}\partial_{y}e_{\lambda}^{b}\omega_{\mu b}^{a}\\
-\frac{1}{4}\partial_{x}\partial_{p}e_{a}^{\nu}e_{\lambda}^{b}\partial_{y}\partial_{p}\omega_{\mu b}^{a}
\end{array}.
\end{equation}
Encouragingly, this matches to first order. Every factor that appears
in (67), (68) and (69) has already been calculated. So even though
expressing (37) in terms of classical index quantities would be tedious,
it would be straightforward.

\medskip{}

\medskip{}
Now that there are expressions for the Christoffel symbols on a deformed
manifold, it is possible to calculate (63). The second order expansion
is:

\begin{equation}
\hat{R}_{\mu\delta\lambda}^{\nu}=R_{\mu\delta\lambda}^{\nu}+i\theta^{\rho\tau}R_{\mu\delta\lambda\rho\tau}^{\nu}+i\theta^{\rho\tau}\theta^{\kappa\sigma}R_{\mu\delta\lambda\rho\tau\kappa\sigma}^{\nu}.
\end{equation}
By performing similar calculations used to reach (61), it can be shown
that that the required coefficients in (70) are

\begin{equation}
R_{\mu\delta\lambda}^{\nu}=\partial_{\delta}\Gamma_{\mu\lambda}^{\nu}-\partial_{\mu}\Gamma_{\delta\lambda}^{\nu}+\Gamma_{\mu\lambda}^{\alpha}\Gamma_{\alpha\delta}^{\nu}+\mu\leftrightarrow\delta
\end{equation}

\begin{equation}
\begin{array}{rl}
R_{\mu\delta\lambda\rho\tau}^{\nu}= & -\partial_{\delta}\Gamma_{\mu\lambda\rho\tau}^{\nu}+\partial_{\mu}\Gamma_{\delta\lambda\rho\tau}^{\nu}-\Gamma_{\mu\lambda}^{\alpha}\Gamma_{\alpha\delta\rho\tau}^{\nu}-\Gamma_{\mu\lambda\rho\tau}^{\alpha}\Gamma_{\alpha\delta}^{\nu}\\
 & +\frac{1}{2}\partial_{\rho}\Gamma_{\mu\nu}^{\alpha}\partial_{\tau}\Gamma_{\alpha\delta}^{\nu}+\mu\leftrightarrow\delta
\end{array}
\end{equation}
and

\begin{equation}
\begin{array}{rl}
R_{\mu\delta\lambda\rho\tau\kappa\sigma}^{\nu}= & \partial_{\delta}\Gamma_{\mu\lambda\rho\tau\kappa\sigma}^{\nu}-\partial_{\mu}\Gamma_{\delta\lambda\rho\tau\kappa\sigma}^{\nu}+\Gamma_{\mu\lambda}^{\alpha}\Gamma_{\alpha\delta\rho\tau\kappa\sigma}^{\nu}\\
 & +\Gamma_{\mu\lambda\rho\tau}^{\alpha}\Gamma_{\alpha\delta\kappa\sigma}^{\nu}+\frac{1}{2}\partial_{\rho}\Gamma_{\mu\nu}^{\alpha}\partial_{\tau}\Gamma_{\alpha\delta\kappa\sigma}^{\nu}\\
 & +\frac{1}{2}\partial_{\rho}\Gamma_{\mu\nu\kappa\sigma}^{\alpha}\partial_{\tau}\Gamma_{\alpha\delta}^{\nu}-\frac{1}{4}\partial_{\rho}\partial_{\kappa}\Gamma_{\mu\lambda}^{\alpha}\partial_{\tau}\partial_{\sigma}\Gamma_{\alpha\delta}^{\nu}+\mu\leftrightarrow\delta.
\end{array}
\end{equation}
Computing the terms in the deformed Riemann tensor is a necessary
step to take, but it must be noted that the Riemann tensor appears
in the Raychaudhuri equation due to the fact that covariant derivatives
do not commute. Up until this point, it has not been made clear what
a covariant derivative would look like if it operated on a manifold
with a minimum length. It is therefore necessary that a prescription
is developed for a covariant derivative which reproduces (35) when
its anti-commutator is calculated.

\medskip{}
\medskip{}
In the classical case,

\begin{equation}
\nabla_{\nu}V^{\rho}=\partial_{\nu}V^{\rho}+\Gamma_{\nu\sigma}^{\rho}V^{\sigma}
\end{equation}
and when a covariant derivative acts on a tensor, a Christoffel symbol
is introduced for each index; positive for up indices and negative
for down indices. I propose that a deformed covariant derivative acts
in almost exactly the same way:

\begin{equation}
\hat{\nabla}_{\nu}V^{\rho}=\partial_{\nu}\hat{V}^{\rho}+\hat{\Gamma}_{\nu\sigma}^{\rho}\star\hat{V}^{\sigma}
\end{equation}
All that is necessary is to keep the partial derivative and use the
deformed Christoffel symbol with a star product instead of regular
multiplication. The true test is to take the star product anti-commutator
and see what results. To redo the standard calculation for the Riemann
tensor, begin with

\begin{equation}
\left[\hat{\nabla}_{\delta}\hat{\nabla}_{\mu}-\hat{\nabla}_{\mu}\hat{\nabla}_{\delta}\right]\star X^{v}
\end{equation}
from which follows

\begin{equation}
\begin{array}{rl}
\left[\hat{\nabla}_{\delta}\hat{\nabla}_{\mu}-\hat{\nabla}_{\mu}\hat{\nabla}_{\delta}\right]\star X^{v}= & \partial_{\delta}\left(\hat{\nabla}_{\mu}X^{v}\right)-\hat{\Gamma}_{\delta\mu}^{\alpha}\star\left(\hat{\nabla}_{\alpha}X^{v}\right)+\hat{\Gamma}_{\delta\sigma}^{v}\star\left(\hat{\nabla}_{\mu}X^{\sigma}\right)\\
 & -\partial_{\mu}\left(\hat{\nabla}_{\delta}X^{v}\right)+\hat{\Gamma}_{\mu\delta}^{\alpha}\star\left(\hat{\nabla}_{\alpha}X^{v}\right)-\hat{\Gamma}_{\mu\sigma}^{v}\star\left(\hat{\nabla}_{\delta}X^{\sigma}\right).
\end{array}
\end{equation}
We should strictly be writing $\hat{X^{v}}$ and $\hat{\Gamma}_{\alpha\sigma}^{v}\star X^{\sigma},$
however it is not necessary to do this because the $X^{v}$ are only
being used to keep track of indices. By expanding (77) and making
use of the distributive property of the $\star$-product, the result
will be

\begin{equation}
\begin{array}{rl}
\left[\hat{\nabla}_{\delta}\hat{\nabla}_{\mu}-\hat{\nabla}_{\mu}\hat{\nabla}_{\delta}\right]\star X^{v}= & \partial_{\delta}\partial_{\mu}X^{v}+\partial_{\delta}\hat{\Gamma}_{\mu\alpha}^{v}X^{\alpha}-\hat{\Gamma}_{\delta\mu}^{\alpha}\star\partial_{\alpha}X^{v}\\
 & -\hat{\Gamma}_{\delta\mu}^{\alpha}\star\hat{\Gamma}_{\alpha\sigma}^{v}X^{\sigma}+\hat{\Gamma}_{\delta\sigma}^{v}\star\partial_{\mu}X^{\sigma}+\hat{\Gamma}_{\delta\sigma}^{v}\star\hat{\Gamma}_{\mu\alpha}^{\sigma}X^{\alpha}\\
 & -\partial_{\mu}\partial_{\delta}X^{v}+-\partial_{\mu}\hat{\Gamma}_{\delta\alpha}^{v}X^{\alpha}+\hat{\Gamma}_{\mu\delta}^{\alpha}\star\partial_{\alpha}X^{v}\\
 & +\hat{\Gamma}_{\mu\delta}^{\alpha}\star\hat{\Gamma}_{\alpha\sigma}^{v}X^{\sigma}-\hat{\Gamma}_{\mu\sigma}^{v}\star\partial_{\delta}X^{\sigma}-\hat{\Gamma}_{\mu\sigma}^{v}\star\hat{\Gamma}_{\delta\alpha}^{\sigma}X^{\alpha}.
\end{array}
\end{equation}
At this point in the classical derivation the no-torsion condition
is assumed. Analogously the assumption that $\hat{\Gamma}_{jk}^{i}=\hat{\Gamma}_{kj}^{i}$
will be made. By doing this, unwanted terms are eliminated from (78)
and the result which appears is

\begin{equation}
\left[\partial_{\delta}\hat{\Gamma}_{\mu\lambda}^{v}-\partial_{\mu}\hat{\Gamma}_{\delta\lambda}^{v}+\hat{\Gamma}_{\delta\sigma}^{v}\star\hat{\Gamma}_{\mu\lambda}^{\sigma}-\hat{\Gamma}_{\mu\sigma}^{v}\star\hat{\Gamma}_{\delta\lambda}^{\sigma}\right]X^{\nu}=\hat{R}_{\lambda\delta\mu}^{v}
\end{equation}
which is exactly what is needed. Since it is possible to show that
the Leibniz rule holds for this new deformed covariant derivative,
all the ingredients to derive a deformed Raychaudhuri equation are
present.

\medskip{}
\medskip{}
Given that the derivation of the Raychaudhuri is known, it is only
necessary to change multiplication to the $\star$-product and add
hats to show that the quantitites which appear are the deformed quantities.

\medskip{}
\medskip{}
There is an important point to make regarding the vectors which appear
in the derivation. Since they are deformed versions of the original
vectors, the substitution $\xi^{a}\rightarrow\hat{\xi^{a}}$ can be
made. At no point will the explicit form of the $\hat{\xi^{a}}$ vectors
be calculated. This is because the vectors will eventually fall out
of the expression for the Einstein Field Equations when the deformed
analogue of Jacobson's derivation is computed.

\medskip{}
\medskip{}
Perform a slightly simplified derivation by writing

\begin{equation}
\hat{B}_{\mu\nu}=\hat{\nabla}_{\nu}\star\hat{\xi_{\mu}}=\partial_{\nu}\hat{\xi_{\mu}}+\hat{\Gamma}_{\nu\mu}^{\alpha}\star\hat{\xi_{\alpha}}
\end{equation}
and define the deformed expansion, shear and twist as

\begin{equation}
\hat{\vartheta}=\hat{B}^{\mu\nu}\star\hat{h}_{\mu\nu},
\end{equation}

\begin{equation}
\hat{\sigma}_{\mu\nu}=\hat{B}_{\left(\mu\nu\right)}-\frac{1}{2}\hat{\vartheta}\star\hat{h}_{\mu\nu}
\end{equation}
and

\begin{equation}
\hat{\omega}_{\mu\nu}=\hat{B}_{\left[\mu\nu\right]}.
\end{equation}
Note that from now on it will be prudent to use the variable $\vartheta$
instead of $\theta$ for the expansion variable in order to avoid
confusing it with the non-commutative parameter. The Raychaudhuri
equation is then derived as

\begin{equation}
\hat{\xi}^{\alpha}\star\hat{\nabla}_{\alpha}\hat{B}_{\mu\nu}=\hat{\xi}^{\alpha}\star\hat{\nabla}_{\nu}\hat{\nabla}_{\alpha}\hat{\xi}_{\mu}+\hat{R}_{\alpha\nu\mu}^{\delta}\star\hat{\xi}^{\alpha}\star\hat{\xi}_{\delta}
\end{equation}

\begin{equation}
=\hat{\nabla}_{\nu}\left(\hat{\xi}^{\alpha}\star\hat{\nabla}_{\alpha}\hat{\xi}_{\mu}\right)-\left(\hat{\nabla}_{\nu}\hat{\xi}^{\alpha}\right)\star\left(\hat{\nabla}_{\alpha}\hat{\xi}_{\mu}\right)+\hat{R}_{\alpha\nu\mu}^{\delta}\star\hat{\xi}^{\alpha}\star\hat{\xi}_{\delta}
\end{equation}

\begin{equation}
=\hat{B}_{\nu}^{\alpha}\star\hat{B}_{\mu\alpha}+\hat{R}_{\alpha\nu\mu}^{\delta}\star\hat{\xi}^{\alpha}\star\hat{\xi}_{\delta}.
\end{equation}
Which allows the trace to be taken, ultimately giving the relation:

\begin{equation}
\frac{d}{d\lambda}\hat{\vartheta}=-\frac{1}{2}\hat{\vartheta}\star\hat{\vartheta}-\hat{\sigma}_{\mu\nu}\star\hat{\sigma}^{\mu\nu}+\hat{\omega}_{\mu\nu}\star\hat{\omega}^{\mu\nu}+\hat{R}_{\alpha\mu}\star\hat{\xi}^{\alpha}\star\hat{\xi}^{\mu}.
\end{equation}

\medskip{}
\medskip{}
Equation (87) is a highly desirable expression. It shows that the
``hats and stars'' prescription carries over to the Raychaudhuri
equation unchanged. Now it is possible to take the deformed Raychaudhuri
equation and use it to repeat Jacobson's calculation. It is not necessary
to repeat the conceptual explanation; so start with

\begin{equation}
\delta Q=-\kappa\int_{H}\lambda\hat{T}_{ab}\star\hat{k}^{a}\star\hat{k}^{b}d\lambda dA
\end{equation}
and feel justified doing this integral, for the reasons outlined in
{[}3{]}. Now make use of the deformed Raychaudhuri equation to get

\begin{equation}
-\kappa\int_{H}\lambda\hat{T}_{ab}\star\hat{k}^{a}\star\hat{k}^{b}d\lambda dA=\int_{H}-\lambda\hat{R}_{ab}\star\hat{k}^{a}\star\hat{k}^{b}d\lambda dA.
\end{equation}
Removing the integrals from (89) gives

\begin{equation}
\hat{T}_{ab}\star\hat{k}^{a}\star\hat{k}^{b}d\lambda dA=\hat{R}_{ab}\star\hat{k}^{a}\star\hat{k}^{b}d\lambda dA.
\end{equation}
and by requiring that the definition for a null surface still holds:

\begin{equation}
\hat{g}_{ab}\star\hat{k}^{a}\star\hat{k}^{b}=0.
\end{equation}
This means that it is possible to recover

\begin{equation}
\hat{T}_{ab}\star\hat{k}^{a}\star\hat{k}^{b}d\lambda dA=\left(\hat{R}_{ab}+\hat{g}_{ab}\right)\star\hat{k}^{a}\star\hat{k}^{b}d\lambda dA
\end{equation}
which becomes

\begin{equation}
\hat{T}_{ab}=\hat{R}_{ab}+\hat{g}_{ab}.
\end{equation}
The Bianchi identities carry through as before, since only indices
are being contracted, which will give the relation:

\begin{equation}
\left(\frac{2\pi}{\hbar\eta}\right)\hat{T}_{ab}=\hat{R}_{ab}-\frac{1}{2}\hat{R}\star\hat{g}_{ab}+\Lambda\hat{g}_{ab}.
\end{equation}
The result in (94) is pleasing and simple, but additional work is
required to expand the expressions to second order in the non-commutative
parameter and to interpret the non-commutative stress-energy tensor.

\medskip{}
\medskip{}
The first issue is not terribly problematic, since it is possible
to work as before and evaluate (94) term-by-term. The Ricci tensor
would be expressed as

\begin{equation}
\begin{array}[t]{rl}
\hat{R}_{\mu\lambda} & =\hat{R}_{\mu\nu\lambda}^{\nu}\\
 & =\hat{R}_{\mu0\lambda}^{0}+\hat{R}_{\mu1\lambda}^{1}+\hat{R}_{\mu2\lambda}^{2}+\hat{R}_{\mu3\lambda,}^{3}
\end{array}
\end{equation}
from which it is possible to calculate the deformed Ricci scalar.
The star product-term would be

\begin{equation}
\hat{R}\star g_{\mu\nu}=\hat{R}\hat{g}_{\mu\nu}+\frac{i}{2}\theta^{ij}\partial_{i}\hat{R}\partial_{j}\hat{g}_{\mu\nu}-\frac{1}{4}\theta^{ij}\theta^{kl}\partial_{i}\partial_{k}\hat{R}\partial_{j}\partial_{l}\hat{g}_{\mu\nu}
\end{equation}
and an expression for the metric can be written as $\hat{g}_{\mu\nu}=\eta_{mn}\hat{e}_{\mu}^{m}\star\hat{e}_{\nu}^{n}$.

\medskip{}
\medskip{}
Additionally, there is a way to deal with a stress-energy tensor living
on a non-commutative space. The idea is to proceed as in {[}6{]} and
say that $\hat{T}_{\mu\nu}$ is the stress-energy tensor of a massive
field living on a non-commutative manifold. If we assume that it is
a massive scalar field which is the gravitational source, then we
can expand $\hat{T}_{\mu\nu}$ in powers of the non-commutative parameter
{[}6{]}:

\begin{equation}
\begin{array}[t]{rl}
\hat{T}_{\mu\nu} & =\frac{1}{2}\left(\partial_{\mu}\phi\star\partial_{\nu}\phi+\partial_{\nu}\phi\star\partial_{\mu}\phi\right)-\frac{1}{2}\eta_{\mu\nu}\left(\partial_{\alpha}\phi\star\partial_{\alpha}\phi-m^{2}\phi\star\phi\right)\\
 & =T_{\mu\nu}+\eta_{\mu\nu}\frac{m^{2}l^{4}}{16}\theta^{\alpha\beta}\theta^{\sigma\rho}\partial_{\alpha}\partial_{\sigma}\phi\partial_{\beta}\partial_{\rho}\phi.
\end{array}
\end{equation}
By doing this, we can take an explicit form of the stress-energy tensor,
calculate the non-commutative corrections and match them, term-by-term,
to the corrections to the geometric variables. This, then, provides
a complete description for calculating a full theory of deformed gravity.

\section{Discussion}

Obtaining non-commutative versions of the Raychaudhuri and Einstein
equations is important for a few reasons. Firstly, they demonstrate
that it is possible to reproduce the Raychaudhuri equation on a noncommutative
spacetime using the Sieberg-Witten map. Although this was only a means
to an end in this paper, a noncommutative Raychaudhuri equation can
be used independently of the study of spacetime thermodynamics.

\medskip{}
\medskip{}
It was stated earlier that putting spacetime on a lattice would break
diffeomorphism invariance and it is legitimate to ask whether imposing
non-commutativity truly does preserve invariance. As was shown in
{[}5{]}, it is possible to use the work of Kontsevich {[}7{]} to retain
the use of the star product, but change its definition to accommodate
for the fact that, under diffeomorphisms, $\theta^{ij}$ becomes a
function of coordinates. This redefining of the star product might
change the appearance of some power series expansions, but more crucially
it shows that it is possible to use the star product and also preserve
diffeomorphism invariance.

\medskip{}
\medskip{}
The derivation of a noncommutative version of the Einstein field equations
is also novel when compared to other attempts to derive noncommutative
spacetime field equations. The normal procedure is to attempt to vary
some action containing a tensor quantity like $\hat{R}$. The variation
of the Einstein-Hilbert action may be straightforward but varying
the noncommutative analogue becomes exceedingly difficult due to the
huge number of terms. It was possible to derive field equations in
this paper with relative ease because Jacobson's approach does not
require the variation of an action.

\medskip{}
\medskip{}
The thermodynamic Van der Waals equation provides a better fit to
reality because it incorporates important physical phenomena that
the ideal gas law neglects. There is strong theoretical justification
for a minimum length in spacetime so it is hoped that by incorporating
this into the Einstein equations that we can match reality ever more
closely. This is clearly desirable, but the fact remains that we are
still doing spacetime thermodynamics. It remains unclear how to use
our current knowledge to go beyond spacetime thermodynamics and start
building a theory of spacetime statistical mechanics.

\medskip{}
\medskip{}
There are important directions to take the work in future. One would
be to start verifying and interpreting solutions to (93). It would
be unsatisfying to simply know that solutions exist - to truly appreciate
the behaviour of non-commutative solutions it would be necessary to
calculate the non-commutative corrections to a few orders. Another
direction to take the work would be to to use Kontsevich's work {[}7{]}
to calculate corrective terms which are guaranteed to not break diffeomorphism
invariance.

\setcounter{page}{21}

\section*{Acknowledgements}
This work was supported by DST/NRF under a South African Research
Chair Initiative Grant. This work formed part of a Masters of Science
in Physics thesis at the University of the Witwatersrand.

\medskip{}
\medskip{}
I am grateful to V. Jejjala for his extensive helpful comments on
the drafts of this paper.

\end{document}